\documentstyle[seceq,epsf,twoside]{ptptex}
\pubinfo{Vol.~{\bf 102}~(2001),~E52-E57}
\notypesetlogo  
\title{${\mib \sigma}$ and ${\mib \kappa}$ in Scattering Processes 
and New ${\mib \pi^0}{\mib \pi^0}$ Phase Shift Data}
\author{%
Kunio {\sc Takamatsu}}
\inst{%
CROSS, Comprehensive Research Organization for Science and Society  \\
 Tsuchiura, Ibaraki 300-0811, Japan. \\
 For the sigma\footnote{
\noindent Sigma Collaboration: M.Y.~Ishida$^{3)}$, S.~Ishida$^{4)}$, 
T.~Ishida$^{5)}$, T.~Komada$^{4)}$, A.M.~Ma$^{1,2)}$, H.~Shimizu$^{6)}$, 
K.~Takamatsu$^{12)}$ and T.~Tsuru$^{5)}$
} and E135\footnote{
\noindent E135 Collaboration: S.~Fukui$^{7)}$, N.~Horikawa$^{8)}$, S.~Inaba$^{5)}$, 
T.~Inagaki$^{5)}$, Y.~Ishizaki$^{9)}$, T.~Iwata$^{8)}$, T.~Kinashi$^{8)}$, 
M.~Kurashina$^{8)}$, I.~Maeda$^{8)}$, T.~Matsuda$^{2)}$, S.~Mori$^{8)}$,
T.~Nakamura$^{10)}$, N.~Nakanishi$^{8)}$, K.~Ohmi$^{5)}$, C.~Oomori$^{5)}$, 
T.~Sato$^{5)}$, K.~Takamatsu$^{2,12)}$, R.~Takashima$^{11)}$, T.~Tsuru$^{5)}$ 
and Y.~Yasu$^{5)}$.\\
1) IHEP, Beijing, 2) Miyazaki U., 3) Tokyo Inst. Tech., 
4) Nihon U., 5) KEK, 6) Yamagata U., 7) Sugiyama W. U., 
8)Nagoya U., 9) INS, 10) U. Kyoto, 11) Kyoto E. U., 12) CROSS.
} collaborations
}
\abst{%
The evidences for $\sigma$(600) and $\kappa$(900) observed in our analyses on 
the $\pi\pi$ and $K\pi$ scattering phase shift data are described, briefly. 
The analysis have been performed by the interfering amplitude method, which 
satisfies the unitarity requirement, using physically meaningful parameters. 
The introduction of the negative phase shifts (repulsive force) are essential 
in the analysis. New data for the $\pi^0\pi^0$ scattering amplitudes and the 
I=0 S wave phase shifts are presented. The data have been obtained in the 
p-p charge exchange reaction, $\pi^- p \rightarrow \pi^0\pi^0 n$ at 9 GeV 
by the E135 experiment at the KEK PS. The amplitude analysis are performed. 
The behavior of the $I$=0 $S$ wave phase shifts below $K\bar{K}$ threshold 
are consistent with those of the $\pi^+\pi^-$, so called, standard data and 
those of the down-flat solution of the CERN-Cracow-M\"unich polarization data.
The analysis of the $\pi^0\pi^0$ phase shift data observes $\sigma$(600) with 
the B-W parameters, $M_\sigma =588 \pm 12$ MeV and 
$\Gamma_\sigma =281\pm 25$ MeV, which are in good agreement with those in 
our analysis on the $\pi^+\pi^-$ data.
}

\begin{document}
\maketitle

\setcounter{tocdepth}{4}

\section{Introduction}

The $\sigma$ particle, the chiral partner of pion as Nambu-Goldstone boson, 
has long been expected for confirmation of its existence in the 
Nambu-Jona-Lasinio type model\cite{rf1}. Though its mass is expected to be 
twice of the quark mass, no $I=0$ scalar meson below 1 GeV had not been 
observed in past two decades, leaving confusions in the studies\cite{rf2}. 
The confusions were due solely to the fact that $\sigma$ was not observed in 
the conventional analysis\cite{rf3} of the $I=0$ $S$ wave $\pi\pi$ scattering 
phase shifts. The phase shift rises slowly up to 90 degrees from the threshold
to 1 GeV and did not allow to accommodate a resonance simply in the region. 
In the review of scalar states, Morgan\cite{rf4} attributes, excluding 
$\sigma$, a very broad resonance around 1GeV to the slowly varying 
$\pi\pi$ phase shift data. Correspondingly, it is also stated clearly that 
the huge events of the $\pi\pi$ system below 1GeV observed in the $pp$ central
collision process should not be recognized as a resonance.

When we presented the result of the $S$ wave resonant state below 1 GeV in the
$\pi^0\pi^0$ system produced in the $pp$ central collision process observed 
by the GAMS spectrometer at CERN\cite{rf5}, Pennington cast a 
comment\cite{rf6} against it on the basis of the "universality argument" 
through the scattering and the production processes. However, the argument 
loses its base, when the right $S$ matrix bases are taken into 
account\cite{rf7}.

Meanwhile, we tried to re-analyze\cite{rf8,rf9} the $\pi\pi$ phase shifts 
data to study $S$ wave resonance below 1 GeV in the scattering process, 
having got the evidence for the sigma particle. Efforts to study the $S$ wave 
resonance in re-analyses of the $\pi\pi$ scattering process have been 
reported by several authors\cite{rf10}, concluding also evidences for 
the existence of $\sigma$

We present in the next chapter a brief summary of the results of our 
reanalysis on the $I=0$ $S$ wave $\pi\pi$ phase shift data to show the 
evidence for the existence of $\sigma$(600) in the scattering process. 
We would also show an evidence for $\kappa$(900) in the results of our 
analysis\cite{rf11} on $K\pi$ scattering phase shift data obtained 
by Aston {\it et al.}\cite{rf12} with the LASS spectrometer at SLAC.
 
The new data on the $\pi^0\pi^0$ system have been obtained in the $\pi^- P$ 
charge exchange process by the E135 experiment using the Benkei 
spectrometer\cite{rf13} at the KEK 12 GeV proton synchrotron. After an 
amplitude analysis of the $\pi^0\pi^0$ system is performed, the  $\pi^0\pi^0$ 
scattering phase shifts and their analysis are described in the 
chapter 3. The study of the  $\pi^0\pi^0$ final state has an advantage that 
the  $\pi^0\pi^0$ scattering amplitude has no odd wave and has no odd isospin 
state. The analyses are not suffered from the huge contribution from $\rho^0$ 
and free from the ambiguity in the phase shift solutions. Though the 
$\pi^0\pi^0$ data have long been desired, the high statistic data by 
Cason {\it et al.}\cite{rf14} are only those used for the analysis in these 
two decades.

\section{${\mib \sigma}$ and ${\mib \kappa}$ in the ${\mib \pi}{\mib \pi}$ and 
${\mib K}{\mib \pi}$ scattering phase shifts.}

The $I=0$ $S$ wave phase shift data of CERN b)\cite{rf15}, 
of Srinivassan {\it et al.}\cite{rf16}, of Rosselet {\it et al.}\cite{rf17} 
and of Bel'kov {\it et al.}\cite{rf18}, so called the standard phase shift 
data, are used in our analysis. The new method, interfering amplitude (IA) 
method\cite{rf8,rf9} is applied for the analysis. It uses a few physically 
meaningful parameters. $S$ matrix is written as 
$S=e^{2i\delta (s)}=1+2i\alpha (s)$, where $\delta (s)$ is the sum of phase 
shifts coming from resonant states, $\delta_R$'s and the background phase 
shift due to a repulsive force below $K\bar{K}$ threshold, 
$\delta_{BG}$, {\it i.e.} $\delta (s)=\delta_R+\delta_{BG}=
\delta_{f_0}+\delta_\sigma+\delta_{BG}$. $f_0$(980) and $\sigma$ are 
considered for the relevant resonant states. The relativistic Breit-Wigner 
form is taken for $\alpha_R (s)$, {\it i.e.} $\alpha_R (s)= 
s\Gamma_R(s)/(m_R^2-s-is\Gamma_R(s))$. Then, the total $S$ matrix is expressed
by the product of $S$ matrices of resonances, $S_R=S_{f_0}S_\sigma$ and that 
of the background, $S_{BG}$, {\it i.e.} $S= S_RS_{BG}$, 
$=S_{f_0}S_\sigma S_{BG}$. It is clear that unitarity for the total $S$ matrix
is automatically satisfied by each $S$ matrix. A hard core type is adopted for
the negative phase shift, $\delta_{BG}$, as $\delta_{BG}=-r_c|{\mib p_1}|$, 
where $r_c$ is a hard core radius and ${\mib p_1}$ is the CM momentum of pion.
The introduction of the negative phases, the repulsive force is not a matter 
of arbitrariness\cite{rf19} but has a physical base. There exists the 
experimental fact that the negative phase shifts\cite{rf20} are observed in 
$I=2$ $S$ wave $\pi\pi$ scattering amplitudes where no resonant state is 
expected. The $I=2$ $S$ wave $\pi\pi$ scattering phase shift, $\delta_0^{(2)}$
decreases linearly from threshold to 1 GeV. It can be expressed by hard core 
parameters\cite{rf9}. The same type of the background phase shifts\cite{rf21} 
may appear in the $I=0$ state. In theoretical consideration\cite{rf21}, the 
repulsive force are derived from the compensating $\lambda\phi^4$ contact 
interaction based on current algebra and PCAC.

\begin{figure}[t]
\parbox{\halftext}{
  \epsfysize=6. cm
  \centerline{\epsffile{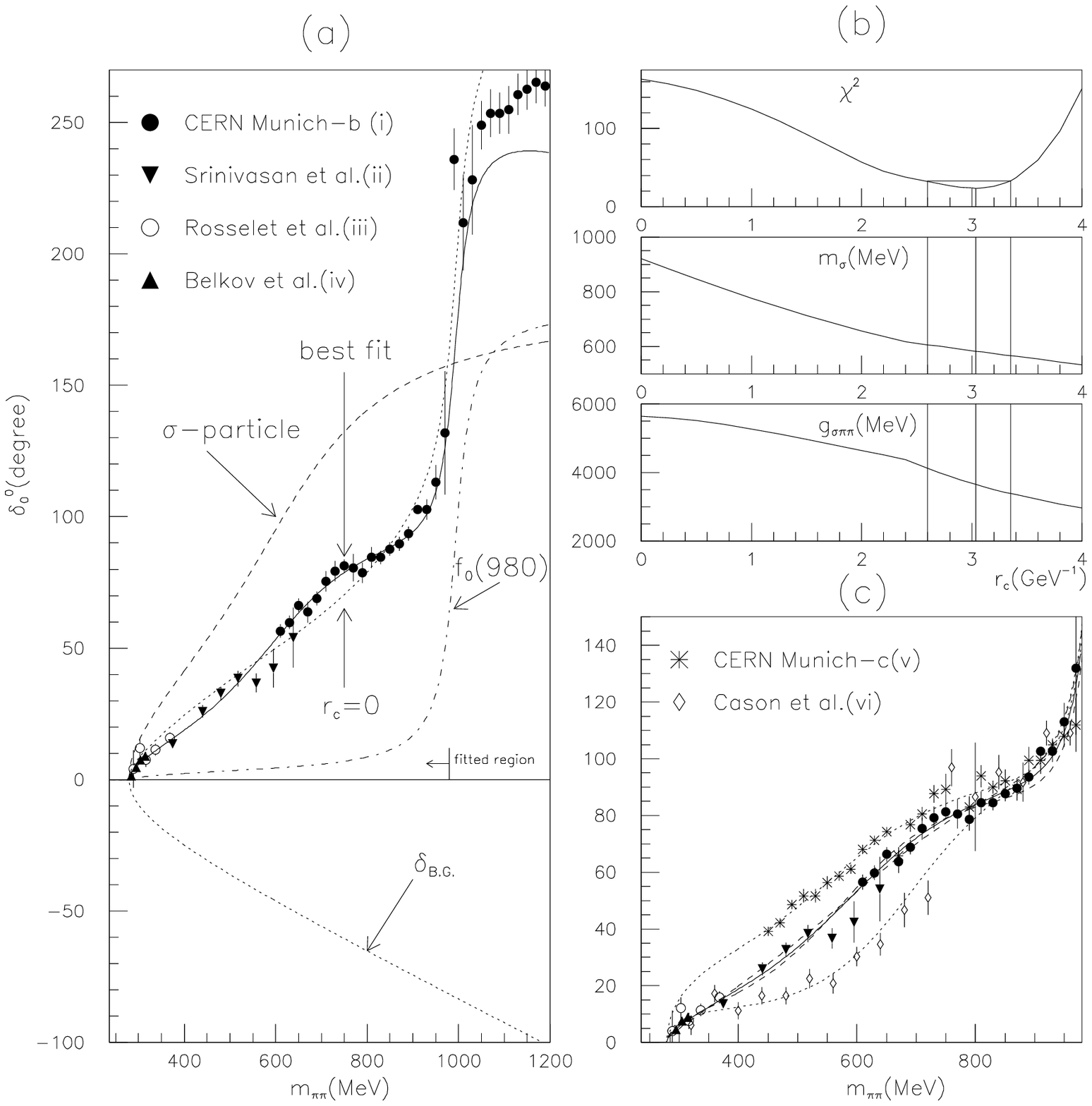}}
  \caption{$I$=0 $\pi\pi$ scattering $S$ wave phase shifts. 
          (a) Best fit to the standard $\delta_0^0$, 
          (b) $\chi^2$, $M_\sigma$ and $g_c$ vs $r_c$, 
          (c) Fittings for uper and lower bounds}
  \label{fig1}}
  \hspace{4mm}
 \parbox{\halftext}{
  \epsfysize=6. cm
  \centerline{\epsffile{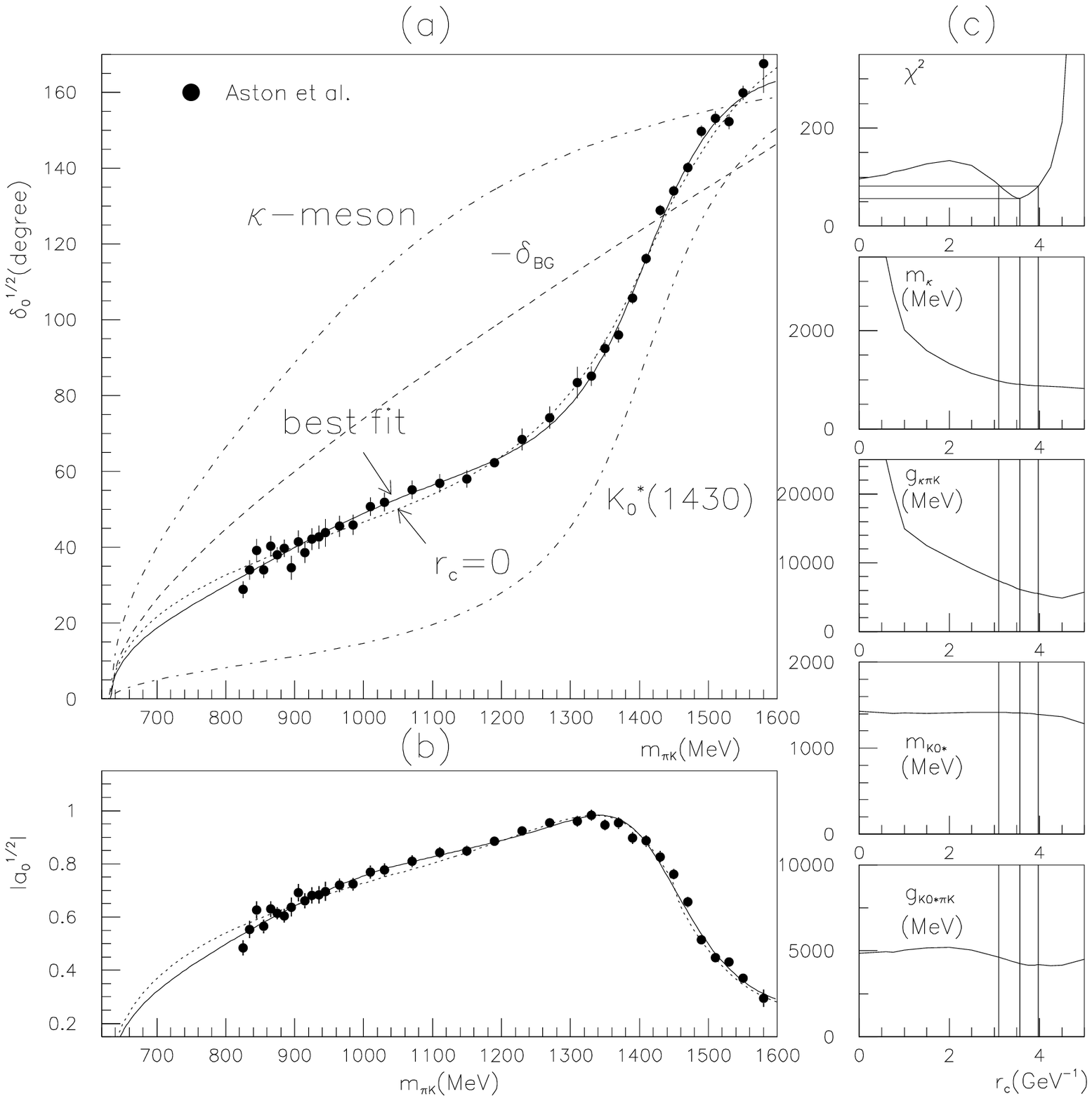}}
  \caption{I=1/2 $K_\pi$ $S$ wave scattering phase shifts.
          (a) Best fit to $\delta_0^{1/2}$, 
          (b) scattering amplitudes,
          (c) $\chi^2$, $M_\kappa$ and $M_{K^{0*}}$ vs $r_c$ 
          }
  \label{fig2}\vspace{0mm}}
\end{figure}
The fitting by the IA method with an introduction of the negative background 
phase shifts reproduces the standard data for $I=0$ $S$ wave scattering phase 
shifts, excellently, as shown by a solid line in Fig.~1(a). The B-W parameters
for $\sigma$ are obtained to be $M_\sigma=585\pm 20$ MeV, 
$\Gamma_\sigma^p=385\pm 70$ MeV and $r_c=3.03\pm 0.35$ GeV$^{-1}$ 
(0.60$\pm$0.07 fm). The reduced $\chi^2$ value is 23.6/(34-4). The existence 
of the low mass $S$ wave resonance with broad width, $\sigma$(600) has been 
confirmed. A comment has been given by Klempt\cite{rf22} on 
our results. The essential role and the physical origin of the negative 
background phase shifts is ignored and our results are looked down upon a 
trivial physics of a $\chi^2$ fitting in the comment.

We analyzed the data of CERN c)\cite{rf23} and of the $\pi^0\pi^0$ scattering 
phase shifts obtained by Cason {\it et al.}\cite{rf14} in the same way that 
for the standard data to get an idea for the upper and the lower bounds of the
parameters in the experimental uncertainties. We obtained 540 MeV and 675 MeV 
for mass values of $\sigma$ and 440 MeV and 345 MeV for widths of it for upper
and lower bounds, respectively. The curves fitted are also shown by dotted 
lines in Fig.1 c).

We have performed a reanalysis of the $K\pi$ scattering phase shift data 
obtained by Aston {\it et al.} in the reaction $K^-P\rightarrow K^-\pi^+n$ 
with the LASS spectrometer at SLAC. The $K^-\pi^+$ scattering amplitudes are 
the sum of the $I$=1/2 and $I$=3/2 components. The $I$=1/2 $S$ wave amplitude
$\delta_0^{(1/2)}$ was determined by subtraction of $I$=3/2 component obtained
by Esterbrook {et al.}\cite{rf24}. The phase shift, $\delta_0^{(1/2)}$ and the
amplitude, $\alpha_0^{(1/2)}$ are shown in Fig. 2 a) and b), respectively. 
The dot-dash lines in the figure a) show the contribution of $\kappa$ meson 
and $K^{0*}$(1430) in the analysis. The negative values of 
$\delta_{BG}^{(1/2)}$ are shown by dotted line in the figure. The parameters 
obtained in the fitting are as follows, 
$M_\kappa=905{{+65}\atop{-30}}$ MeV, $\Gamma_\kappa=545{{+235}\atop{-110}}$ 
MeV and $r_{oc}^{(1/2)} = 3.57 {{+0.40}\atop{-0.45}}$ GeV$^{-1}$ 
(0.7 ${{+0.08}\atop{-0.09}}$ fm). The existence of $\kappa$(900) has been 
confirmed. It is interesting in the SU(3) flavor symmetry point of view to 
note that value of the core radius is almost the same that for I=0 $S$ wave 
$\pi\pi$ phase shifts.

\section{New data for the ${\mib \pi^0}{\mib \pi^0}$ system}

\begin{figure}[t]
\parbox{\halftext}{
  \epsfysize=8. cm
  \centerline{\epsffile{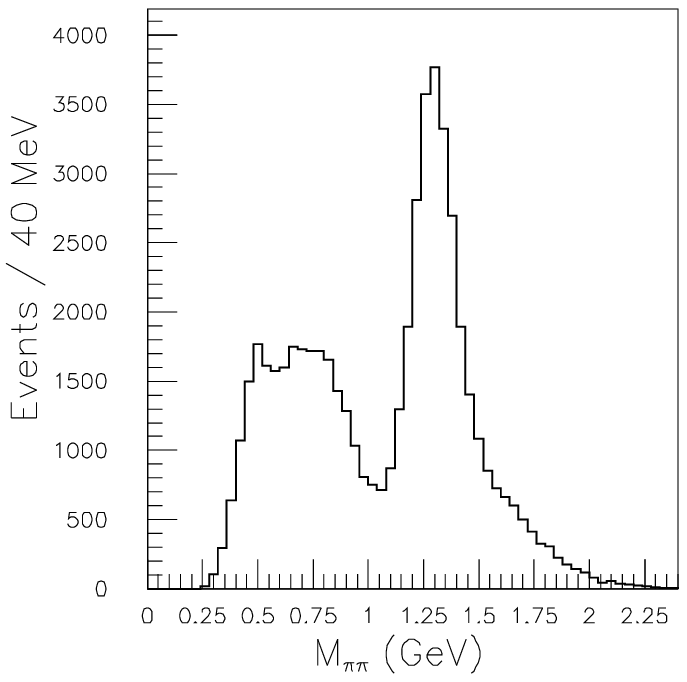}}
  \caption{Acceptance corrected $\pi^0\pi^0$ mass distribution.}
  \label{fig3}}
 \hspace{4mm}
 \parbox{\halftext}{
  \epsfysize=8. cm
  \centerline{\epsffile{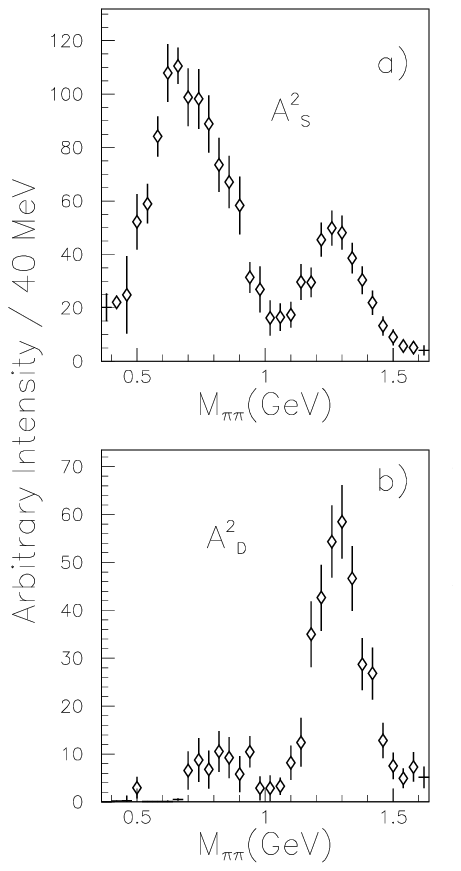}}
  \caption{Results of the partial wave analysis. a) 
   for $S$ wave and b) for $D$ wave.}
  \label{fig4}\vspace{0mm}}
\end{figure}
The phase shift data for the $\pi^0\pi^0$ system has long been desired. 
However, the experimental difficulties and probably the strong argument on no 
existence of $\sigma$ in past two decades suppressed efforts to get data of 
the $\pi^0\pi^0$ scattering phase shifts with enough quality for the analysis.
The high statistic data of Cason {\it et al.}\cite{rf14} have only been used 
for analysis, so far. The data, however, show different behavior, apparently, 
from the $I$=0 $S$ wave $\pi^+\pi^-$ phase shift data as seen in Fig.~1 c).
We have analyzed the data of the $\pi^0\pi^0$ final state produced in the 
$\pi^-P$ charge exchange process, $\pi^-P \rightarrow \pi^0\pi^0 n$ at 9 GeV 
studied by E135 at the KEK 12 GeV proton synchrotron. We have obtained the 
$\pi^0\pi^0$ scattering amplitudes up to 1.5 GeV and the 
$\pi^0\pi^0$ scattering phase shifts below $K\bar{K}$ threshold. Data was 
taken by Benkei Spectrometer late in 1980's. The spectrometer system is 
described elsewhere\cite{rf13}. We used the data of the all neutral final 
state (four $\gamma$'s) for analysis. The acceptance corrected mass 
distribution of the $\pi^0\pi^0$ (reconstructed from four $\gamma$'s) system 
is shown in Fig. 3.

The off mass-shell scattering amplitude, 
$T_{\pi\pi}(m_{\pi\pi}^2, cos\theta, t)$ are extrapolated to 
the on mass-shell scattering amplitude at the pion pole, 
$T_{\pi\pi}(m_{\pi\pi}^2, cos\theta, m_\pi^2)$, as the
process can be considered to proceed through one pion exchange. 
A linear extrapolation is adopted. The on mass-shell scattering 
amplitude can be described by the $S$ and $D$ waves taken in consideration, 
as follows, 
$T_{\pi\pi}(m_{\pi\pi}^2, cos\theta, m_\pi^2)
=A_S +A_D 5(3cos2\theta -1)/2$, 
where $A_S$ and $A_D$ are $S$ and $D$ wave scattering amplitudes, 
respectively. The partial waves for the $S$ and $D$ waves, obtained are shown 
in Fig.~4 a) and b) respectively.
 
The $S$ wave $\pi^0\pi^0$ scattering amplitudes can be written in terms of 
the of $I=0$ $S$ wave scattering phase shift, $\delta_0^{(0)}$ and of the 
$I$=2, $\delta_0^{(2)}$, $|A_s|^2\sim sin^2(\delta_0^{(0)}-\delta_0^{(2)})$ 
below $K\bar{K}$ threshold. We used a hard core type for $\delta_0^{(2)}$, 
as $\delta_0^{(2)}=-r_c^{(2)}|{\mib q_1}|$, where ${\mib q_1}$ is the CM 
momentum of pion and $r_c$ is the core radius. The parameter, $r_c^{(2)}$ 
has been obtained\cite{rf25} from $\pi^+\pi^-$ data to be 0.87 GeV$^{-1}$ 
(0.17 fm). 
The $S$ wave $\pi^0\pi^0$ phase shifts obtained below $K\bar{K}$ threshold 
are shown in Fig.~5 with open squares. The results are consistent with 
so called the standard data of the $\pi^+\pi^-$ phase shifts and also with 
those of the down-flat solution\cite{rf26} obtained in the analysis performed 
recently by Kami\'nski {\it et al.} on the CERN-Cracow-M\"unich polarization 
data\cite{rf27}.

The phase shift difference $\delta_0^{(0)}-\delta_0^{(2)}$ at the neutral 
kaon mass in our analysis gives 42.5$\pm$5 degrees, which is consistent with 
the value obtained from the threshold pion production\cite{rf28}.

\begin{figure}[t]
  \epsfysize=6 cm
 \centerline{\epsffile{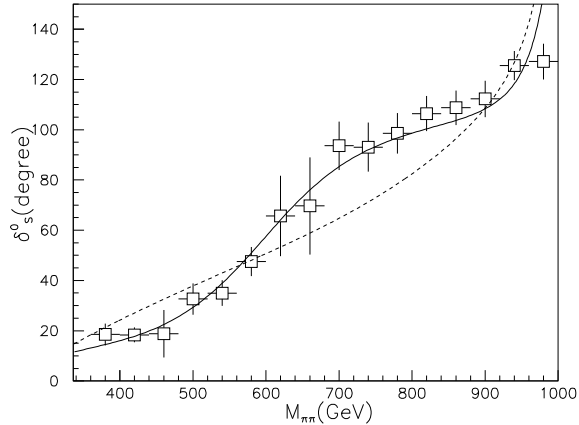}}
 \caption{$I$=0 $\pi^0\pi^0$ $S$ wave scattering phase shifts. The solid
 and the dotted lines show the fits with the hard core and without hard core,
 respectively.}
  \label{fig5}
\end{figure}
The $\pi^0\pi^0$ phase shift data are analyzed by the IA method. Negative 
background phase shifts are introduced in the analysis. A hard core is used 
for them. The fit is shown in Fig.~5 by the solid line. The Breit-Wigner 
parameters obtained for $\sigma$ are as
follows; $M_\sigma=588\pm 12$ MeV, $\Gamma_\sigma$=281$\pm$25 MeV 
and $r_c$=2.76$\pm$0.15 GeV$^{-1}$. The reduced $\chi^2$ value is 20.4/12. 
These values are in good agreement with those which we have obtained in
our analysis\cite{rf8,rf9} on the standard $\pi^+\pi^-$ phase shift data. 
The dotted line in the figure is the result obtained with no negative 
background (no hard core, $r_c$=0). The B-W parameters obtained are 
$M_{``\sigma"}=890\pm 16$ MeV, and $\Gamma_{"\sigma"}=618\pm51$ MeV which 
deviate appreciably from those with the hard core. The reduced $\chi^2$ 
becomes worse to be 85.0/13.

\vspace{-0.25em}
\section{Conclusions}
\vspace{-0.25em}

Evidences for $\sigma$(600) and $\kappa$(900) in our analyses on the 
scattering processes are summarized. The existence of them suggests strongly 
that they form a chiral nonet. The new data for the $\pi^0\pi^0$ scattering 
amplitudes are presented for the $\pi^-P$ charge exchange reaction at 9 GeV/c.
The $\pi^0\pi^0$ $S$ wave scattering phase shifts below 1 GeV show the 
consistent behavior with the $\pi^+\pi^-$ standard phase shift data and with 
those of the CERN Cracow M\"unich polarization data. $\sigma$(600) is 
observed in the analysis of the $\pi^0\pi^0$ scattering phase shifts. Its 
Breit-Wigner parameters are 588 MeV and 276 Mev for mass and width, 
respectively, which are in excellent agreement with those for the 
$\pi^+\pi^-$ phase shift data.


\begin{thebibliography}{99}
\bibitem{rf20}Data are summarized in the Fig.2 of ref.9).
\end{thebibliography}
\end{document}